\documentclass[aps,pra,notitlepage,onecolumn,superscriptaddress]{revtex4-2}

\usepackage[tmargin=1in, bmargin=1in, lmargin=1.25in, rmargin=1.25in]{geometry}
\usepackage{bbm}
\usepackage{mathrsfs}
\usepackage{epsfig}
\usepackage{soul,xcolor}
\usepackage{graphicx}
\usepackage{amsfonts}
\usepackage{amsthm}
\usepackage[figuresright]{rotating}
\usepackage{amssymb}
\usepackage{amsmath}
\usepackage[shortlabels]{enumitem}
\usepackage{mathtools}
\usepackage{dcolumn}
\usepackage{physics}
\usepackage{float}
\usepackage{bm}
\usepackage{verbatim}
\usepackage{thm-restate}
\usepackage{setspace}
\usepackage[colorlinks,linkcolor=blue,anchorcolor=blue,citecolor=blue,urlcolor=blue]{hyperref}
\usepackage{stackengine}

\DeclarePairedDelimiter{\bracks}{[}{]}
\DeclarePairedDelimiter{\parens}{(}{)}

\newcommand{\R}{\mathbb{R}}
\newcommand{\Z}{\mathbb{Z}}
\newcommand{\nand}{\textsc{nand}}

\usepackage[capitalise]{cleveref}
\theoremstyle{plain}
\newtheorem{theorem}{Theorem}

\theoremstyle{definition}
\newtheorem{definition}[theorem]{Definition}
\newtheorem{remark}[theorem]{Remark}

\begin{document}

\title{Resource savings from fault-tolerant circuit design}
\author{Andrew K. Tan}
  \email{aktan@mit.edu}
  \affiliation{Department of Physics, Co-Design Center for Quantum Advantage, Massachusetts Institute of Technology, Cambridge, Massachusetts 02139, USA}
\author{Isaac L. Chuang}
  \affiliation{Department of Physics, Co-Design Center for Quantum Advantage, Massachusetts Institute of Technology, Cambridge, Massachusetts 02139, USA}
  \affiliation{Department of Electrical Engineering and Computer Science, Massachusetts Institute of Technology, Cambridge, Massachusetts 02139, USA}

% \vspace*{-.6in}

\begin{abstract}
  Using fault-tolerant constructions, computations performed with unreliable components can simulate their noiseless counterparts though the introduction of a modest amount of redundancy.
  Given the modest overhead required to achieve fault-tolerance, and the fact that increasing the reliability of basic components often comes at a cost, are there situations where fault-tolerance may be more economical?
  We present a general framework to  account for this overhead cost in order to effectively compare fault-tolerant to non-fault-tolerant approaches for computation, in the limit of small logical error rates.
  Using this detailed accounting, we determine explicit boundaries at which fault-tolerant designs become more efficient than designs that achieve comparable reliability through direct consumption of resources.
  We find that the fault-tolerant construction is always preferred in the limit of high reliability in cases where the resources required to construct a basic unit grows faster than \(\log(1 / \epsilon)\) asymptotically for small \(\epsilon\).
\end{abstract}

\maketitle

%%%%%%%%%%%%%%%%%%%%%%%%%%%%%%%%%%%%%%%%%%%%%%%%%%%%%%%%%%%%%%%%%%%%%%%%%%%%%%%%
\section{Introduction}
  Complex computations can be built from simple primitives, but what if these primitive operations are themselves noisy?
  One may imagine that the errors accumulate with each step, effectively limiting the complexity of computations that can be derived from simple primitives.
  Remarkably, this is not the case: using a biologically-inspired framing of the problem, the fact that noisy systems could simulate noiseless computations using modest redundancy was first discovered by von Neumann in 1952 \cite{von-neumann1956probabilistic}.
  The key observation was that the probability of error can be bounded by a constant in a manner independent of computation depth by encoding data in an error correcting code and interleaving computation with error correction:
  this is the \emph{von Neumann} construction for fault-tolerance.

  In the decades since von Neumann's pioneering work, fault-tolerance has been studied using the formalism of the Boolean circuit, where the primitives are Boolean logic gates and computation are described by a directed acyclic graph of dependencies.
  von Neumann's construction, formalized using a Boolean circuit model of computation has since been constructively sharpened for formulas, i.e. circuits with fan-out \(1\), with results conjectured to hold more generally \cite{hajek1991on-the-maximum,evans1998on-the-maximum,evans2003on-the-maximum}.
  More recent work has also extended positive results to circuit models with larger alphabet sizes \cite{tan2023on-reliable}, neural network models of computation \cite{zlokapa2022biological}.  Fault tolerance is also vital to the promise of large-scale quantum computation \cite{shor1995scheme,preskill1998fault-tolerant,campbell2017roads}.
  On the other hand, a number of negative results effectively place bounds on both fault-tolerance thresholds and overhead requirements \cite{feder1989reliable,evans1999signal,unger2007noise,unger2010better}.

  \emph{Motivation and related work---}
  While fault-tolerance is an interesting theoretical exercise, one may wonder if it is of practical relevance: why bother with the additional complexity required to design a fault-tolerant circuit, when one can simply build a more perfect gate?
  It is true that for standard transistor-based logic gates have such low physical error rates \cite{schroeder2009dram} that one is often better served using lightweight error detection methods at the software-level rather than implementing full hardware-level fault-tolerance.
  However, for low-power and nano-scale semiconductor devices, we are approaching a point where increased noise and statistical variations in manufacturing have lead some to call for new, more robust, computational paradigms \cite{shanbhag2008the-search,shanbhag2018shannon-inspired}.
  Inspiringly, Chatterjee and Varshney \cite{chatterjee2016energy-reliability,chatterjee2020energy-reliability} applied the negative fault-tolerance results of \cite{evans1999signal} to place bounds on the energy-reliability tradeoffs allowed for nano-scale circuits and deep feed-forward neural networks.
  Their approach provides insightful scaling results, but to make these ideas useful for practical computational systems, the theory of fault-tolerance bounds must become constructive.

  Suppose that instead of perfect computation, the goal is to offer some specific level of reliability.  This is increasingly a desirable systems engineering goal for computation, particularly when algorithmic outputs are probabilistic or the problem is inherently non-deterministic, e.g. as often is the case in machine learning.
  Fault-tolerance constructions can offer such an engineering tradeoff: 
  by increasing the size of the code, a computation can be performed by a noisy computer to arbitrary precision using polylogarithmic overhead in the number of gates \cite{von-neumann1956probabilistic,pippenger1988reliable,feder1989reliable,evans1998on-the-maximum,evans1999signal,nielsen2010quantum}.  
  Importantly, fault-tolerance need not just be used to obtain a vanishingly small error rate; the desired error rate can be dialed in by changing the amount of redundancy employed.  
  And moreover, the resource cost for fault-tolerance may come in many forms, not just the energy consumed, but also the space or time required.  
  Thus, given the in-principle relatively modest, (poly)logarithmic, overhead required by fault-tolerant designs \cite{von-neumann1956probabilistic,nielsen2010quantum}, it seems natural to wonder if there are constructive approaches to show whether fault-tolerance may actually provide net savings for a broad class of resource--reliability trade-offs.

  Thaker et al. \cite{thaker2005recursive,thaker2008on-using} showed that in principle fault-tolerant constructions based on \emph{recursive triple modular redundancy} may be more resource efficient than their non-fault-tolerant counterparts.
  Impens \cite{impens2004fine-grained} studied the same recursive construction as a way to trade resources for reliability, i.e. reliability as a fungible resource, and showed that fault-tolerant constructions may be more resource-efficient than their non-fault-tolerant counterparts if the computational primitive follows certain reliability--resource trade-offs.
  While similar to this work in spirit, both Thaker et al. and Impens use a recursive concatenation-based fault-tolerant design in which overhead scaling is polylogarithmic and constants are difficult to obtain owing to the fact that the size of the resulting fault-tolerant circuits grow exponentially with the level of recursion.

  Our approach builds on this body of prior work, and goes further by employing a constructive fault-tolerant procedure which allows precise overhead estimates with no unbounded constants. 
  Our construction allows the overhead to be separated into three main components: the first contribution is due to the code size requirement to attain the desired level of certainty, effectively a concentration bound;
  the second contribution is from the number of gates required to the error correcting circuit; and finally,
  a third contribution comes from the statistical dependence in a circuit's outputs, which depends on the details of the circuit's connectivity. 
  We use a fault-tolerant construction based on fixed-depth error correction circuits in which overhead is logarithmic, thus simplifying the analysis compared with the recursive construction used by Thaker et al. and Impens.
  Our fixed-depth model allows us to numerically estimate constant factors required to determine the point at which fault-tolerant designs become more efficient than designs that achieve comparable reliability through direct consumption of resources.

  \emph{Roadmap---}
  The rest of the paper is organized as follows.
  First, we formalize the notion of a constant-depth fault-tolerant construction.
  Focusing on the simplest, depth-\(2\) construction, we perform a careful analysis of the asymptotic overhead in the number of gates (\cref{sec:gate-overhead}).
  We find an overhead that is logarithmic in the overhead of desired reliability and numerically determine the constant factors.
  This is followed by the introduction of the asymptotic resource--reliability trade-off (\cref{sec:resource-overhead}).
  Here we find three qualitatively different cases and determine the regions of design space in which the fault-tolerant construction may be more resource efficient.
  Finally, we discuss potential avenues to further improve the fault-tolerant overhead and speculate on abstract models of computation where our results may provide insights (\cref{sec:concluding-remarks}).

%%%%%%%%%%%%%%%%%%%%%%%%%%%%%%%%%%%%%%%%%%%%%%%%%%%%%%%%%%%%%%%%%%%%%%%%%%%%%%%%
\section{Fault-tolerance number overhead}
\label{sec:gate-overhead}
  First, we review a few basics of circuit-based fault-tolerance.
  In the most commonly studied scenario \cite{von-neumann1956probabilistic}, we are given a set of Boolean gates subject to some noise---these are the \emph{basic units} of our computation.
  For our purposes, we will assume that a noisy gate behaves as an ideal gate save for a probability \(\epsilon_{\text{P}}\) that its output is flipped; for simplicity, we assume additionally that all basic gates are subject to the same level of noise \(\epsilon_{\text{P}}\), and that the noise is independent.
  The goal is to build a \emph{logical unit} using a number of basic gates which acts on encoded data such that its errs with error probability \(\epsilon_{\text{L}}\).
  Exactly how such a fault-tolerant logical unit may be constructed, and exactly what it means for it to make an error will be discussed later.
  What is important, however, is that in addition to the overhead introduced by building these logical units, these logical units only remain reliable up for \(\epsilon_{\text{P}} < \epsilon^*\) where \(\epsilon^*\) is a threshold.
  Note that while we call \(\epsilon^*\) the threshold, for our fault-tolerant constructions \(\epsilon^*\) may more accurately be described a pseudothreshold, with the threshold being reserved for the limiting pseudothreshold \cite{svore2005a-flow-map}. 
  To summarize our notation:
  \begin{itemize}[noitemsep,topsep=0pt]
    \item \(\epsilon_{\text{P}}\) is the physical error rate of a basic unit;
    \item \(\epsilon_{\text{L}}\) is the error rate of a logical unit; and,
    \item \(\epsilon^*\) is the fault-tolerance pseudothreshold.
  \end{itemize}
  In general, we have \(\epsilon_{\text{L}} \ll \epsilon_{\text{P}} < \epsilon^*\).

  There are two ways one might go about designing a circuit using faulty components to achieve a target logical error rate \(\epsilon_{\text{L}}\) in light of the existence of fault-tolerant constructions.
  In the first, i.e. the non-fault-tolerant route, we may simply elect to work with higher fidelity gates, choosing physical error rate \(\epsilon_{\text{P}} = \epsilon_{\text{L}}\).  
  Alternatively, we may use a fault-tolerant construction, allowing us to choose to operate with gates at an intermediate error rate \(\epsilon_{\text{P}} \in (\epsilon_{\text{L}}, \epsilon^*)\), where \(\epsilon^*\) is the pseudothreshold of our fault-tolerant construction.  The resource cost of the two routes is generally very different, and our goal will be to accurately model these costs so that the two routes may be compared quantitatively.

  Next, we formalize the concept a fault-tolerant circuit construction.
  For the purposes of our discussion we make the following definition:
  \begin{definition}[Circuit]
  \label{def:circuit}
    A \emph{circuit} is a triple \(\mathcal{C} = (G, L, K, F)\) where
    \begin{itemize}[noitemsep,topsep=0pt]
      \item \(L\) is a set of labels \(\ell\);
      \item \(K\) is a set of positive integers indexed by elements of \(L\), \(k_\ell \in \Z_+\); 
      \item \(G\) is a directed acyclic graph with vertices \(V\) and edges \(E\) where each vertex is associated with a label \(\ell \in L\), and furthermore each vertex with label \(\ell\) has either in-degree \(k_\ell\) or in-degree \(0\); and,
      \item \(F\) is a set of Boolean functions indexed by elements of \(L\), \(F_{\ell}: \{0, 1\}^{k_\ell} \rightarrow \{0, 1\}\).
    \end{itemize}
    Vertices with in-degree \(0\) \emph{inputs}, and those with out-degree \(0\) are called \emph{outputs}.
  \end{definition}
  We can associate with each label \(\ell \in L\) in our definition with a type of gate with specified fan-in \(k_\ell\);
  each vertex in graph \(G\) then represents a gate with edges describing connectivity.
  Vertices with in-degree \(0\) are assumed to take a number of input wires dependent on the gate type and vertices with out-degree \(0\) are gates that provide a single output bit for each assignment of input wires. 
 
  Using our formal definition of a circuit, we turn to formalizing the fault-tolerant construction.
  \begin{definition}[Fault-tolerant circuit construction]
  \label{def:ft-construction}
    Let \(\mathcal{C} = (G, L, K, F)\) be a circuit.
    Further let \(\mathcal{R}_n = (e_n ,d_n)\) for \(n \in \Z_+\) be a \([n, 1]\) error correcting block with encoder \(e_n: \{0, 1\} \rightarrow \{0, 1\}^n\) and decoder \(d_n: \{0, 1\}^n \rightarrow \{0, 1\}\).
    For a fixed gate set, a \emph{fault-tolerant construction} \(\mathcal{FT}_n\) over code \(\mathcal{R}_n\) is specified by a family of operators that maps each vertex \(v\) of type \(\ell\) in the graph \(G\) to a circuit \(C_v\), i.e. \(\operatorname{FT}^{(\ell)}_n: v \mapsto \mathcal{C}_v\), for \(n \in \Z_+\) and all \(\ell \in L\) satisfying the following properties:
    \begin{enumerate}[i),noitemsep,topsep=0pt]
      \item The circuit \(\operatorname{F}^{(\ell)}_n(v) = C_v\) can be written \(C_v = (G_v, L, K, F)\), i.e. using the set of gates as \(\mathcal{C}\);
      \item \(C_v\) has exactly \(k_\ell\) \emph{bundles} of inputs each of size \(n\), and one size-\(n\) bundle output; 
      \item for all \(n\), the depth of circuit \(C_v\) is bounded by some constant \(D \in \Z_+\); 
      \item if all signals within the input bundles are set to \(e_n(x_i)\), \(i \in \{1, \dots, k_\ell\}\), the outputs of \(C_v\) satisfies truth table \(F_\ell \circ d_n\); 
      \item if input signals are subject to noise of strength \(\Delta\) and the outputs of each gate are subject to noise \(\epsilon_{\text{P}}\), the output bundle \(C_v\) must amplify the signal for some non-empty range of \(\Delta\) up to some pseudothreshold \(\epsilon^* > 0\) for all \(n > n_0 \in \Z_+\) (amplification formally defined in \cite{shutty2020tight}); and,
      \item in this amplification region, the probability of a logical error \(\epsilon_{\text{L}}\) (i.e. a mismatch between the decoded output and desired truth table) as \(n \rightarrow \infty\) satisfies \(\epsilon_{\text{L}} \sim e^{-\theta(n)}\).
    \end{enumerate}
    Given the family of operations \(F_n\), one may derive an equivalent fault-tolerant by replacing each vertex using \(\operatorname{FT}_n\) and connecting input and output bundles according to \(G\).
    In general, a randomizing permutation may need to be applied to input or output bundles in order to minimize dependence between wires when gates are applied in sequence.

    We denote by \(\mathcal{FT}_n(\mathcal{C})\) the fault-tolerant circuit derived from \(\mathcal{C}\) using this procedure.
  \end{definition}
  Note that requirement \cref{def:ft-construction}iii excludes the concatenation-based approaches to fault-tolerance employed, for example in \cite{nielsen2010quantum,impens2004fine-grained,svore2005a-flow-map,thaker2005recursive}.  This is being deliberately done, in order to separate the width and depth components of the fault-tolerant construction discussed in \cref{ssec:scaling-arguments}.
  In the nomenclature of this paper, each level of the concatenation-based approach is equivalent to the choice of a different error correction architecture (see \cref{rem:larger-error-correcting-circuits}).

  Also, while \cref{def:ft-construction} allows data to be encoded in a general error correcting code, for the purposes of our discussion, we will consider following fault-tolerant construction based on the \([n, 1]\) repetition code and depth-\(2\) majority circuit for error correction:
  \begin{remark}[Depth-\(2\) fault-tolerant construction]
  \label{rem:depth-2-ft-construction}
    Consider the construction studied by \cite{von-neumann1956probabilistic,evans1998on-the-maximum,unger2007noise} using \(2\)-input \nand~gates, i.e. \(L = \{\nand\}\) and \(k_\nand = 2\).
    Our operator \(\operatorname{FT}^{(\nand)}_n\) replaces each \nand~gate in the original circuit with \(n\) \nand~gates.
    These \(n\) \nand~gates are then followed by \(2 n\) gates used for error correction.
    If we label each gate by a tuple \((l, i)\) for layer \(l \in \{1, 2\}\) and index \(i \in \{0, \dots, n-1\}\), the construction connects \((l, i)\) to the output of gates \((l - 1, i)\) and \((l - 1, i - 1 \mod{n})\);
    where \((0, i)\) denotes the \(i\textsuperscript{th}\) input wire.
    The error correction component of this construction is depicted in \cref{fig:bayesian-network}.
  \end{remark}
  We discuss the properties of depth-\(2\) construction of \cref{rem:depth-2-ft-construction} in \cref{ssec:ft-formulas,ssec:ft-circuits}.

  Equipped with \(\mathcal{FT}_n\), we are able to convert any circuit into an equivalent fault-tolerant circuit. The subject of the remainder of this paper will be a careful accounting of the resources required to target a logical error rate using this construction.
  Assuming that the resource utilization of a circuit is proportional to the number of basic units of which the circuit is comprised, the number of basic units required by the fault-tolerant circuit is of central importance.

  We define the \emph{number overhead} (or equivalently the \emph{gate overhead} for circuit-based models of computation) to be the number of additional basic components required by \(\mathcal{FT}_n(\mathcal{C})\) when compared with \(\mathcal{C}\).
  For the remainder of this Section, we turn our focus to the asymptotics of this number overhead in terms of the number of basic units required to achieve a desired logical error rate, specifically the constants in the exponent of \cref{def:ft-construction}vi.
  The \emph{resource overhead} required for fault-tolerance can be calculated given the number overhead and the resource--reliability trade-off for the basic unit and will be the subject of \cref{sec:resource-overhead}.

\subsection{Scaling arguments}
\label{ssec:scaling-arguments}
  \begin{figure}
    \includegraphics[width=0.95\textwidth]{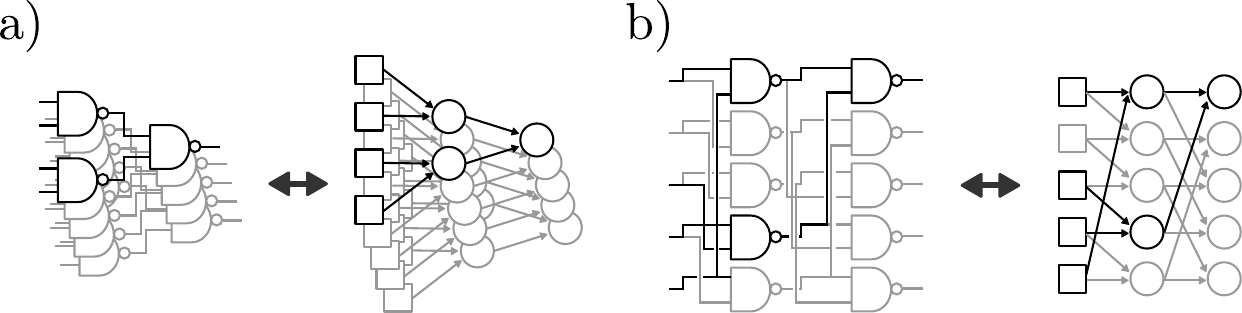}
    \caption{Error correcting depth-\(2\) majority a) formula, and b) an equivalent circuit for the repetition code of size \(n = 5\).
    In both cases, the circuit is shown on the left, with the induced Bayesian network depicted on the right (squares representing input wires, and circles representing computed wires). 
    The computation path of one output wire is highlighted in black throughout with the rest of the circuit in grey.
    }
  \label{fig:bayesian-network}
  \end{figure}

  For our analysis, we separate the overhead into three components.
  The first is due to the required code size, i.e. its `width' overhead, which depends on the target error rate \(\epsilon_{\text{L}}\), and the error rate of the individual output wires \(\epsilon\).
  The second, the `depth' overhead, is related to the fraction of layers dedicated to error correction and is dependent on the operating operating point \(\epsilon_{\text{P}}\) and a fiducial error rate \(\Delta \in (0, 1/2)\); 
  in other words, if \(r\) denotes the number of wires in the repetition code of size \(n\) in the \(1\) state, a logical \(1\) is encoded if \(r > n(1 - \Delta)\), a logical \(0\) is encoded if \(r < n \Delta\), and otherwise the signal is considered erroneous.
  In the fault-tolerant regime, we have that the probability of error in any wire can be maintained \(\Delta < \frac{1}{2} - \Omega(1)\) independently of circuit depth.
  The final component is due to an effective reduction of code size due to statistical dependencies in the output of circuits which is absent in the case of formulas.

  \emph{Width overhead---}
  A key property of fault-tolerance is that a constant, depth-independent logical error rate is maintained by interleaving computation with layers dedicated to error correction.
  In order to accomplish this, the data must be encoded in an error correcting code throughout the computation.
  Increasing the size of this code is the key to reducing logical errors and is the width overhead.
  Since the logical error rate is suppressed by \(e^{\Theta(d)}\), where \(d\) is the distance of the code.
  From a simple concentration bound, we find that it suffices to choose a code of distance
  \begin{equation}
  \label{eq:d-asymptotics}
    d(\epsilon, \epsilon_{\text{L}}) = \Theta\parens*{\log\parens*{\frac{1}{\epsilon_{\text{L}}}}}.
  \end{equation}
  Assuming the use of a linear-distance code, this implies a width overhead of \(\Theta\parens*{\log\parens*{1 / \epsilon_{\text{L}}}}\).
  In the fault-tolerance setting, we must account both for errors introduced during computation and in the process of error correction and thus the hidden constant factor in \cref{eq:d-asymptotics} is in general also dependent on \(\epsilon_{\text{P}}\) and \(\Delta\).
  In general, we write the width overhead as \(n(\epsilon_{\text{L}}; \epsilon_{\text{P}}, \Delta)\).

  One may object to the fact that in the fault-tolerant construction, the inputs and outputs of the computation are encoded in a bundles of \(n\) wires, and not strictly comparable to that of the non-fault-tolerant construction.
  While this is true, we may assume that we have access to a perfect (or otherwise highly reliable) encoder and decoder which we may invoke respectively at the beginning and end of the computation.
  The use of these idealized encoders and decoders is simply a constant cost which contributes negligibly to the overall resource cost in the limit of long computations, which is the primary concern of fault-tolerant constructions.

  \emph{Depth overhead---}
  For any error correction construction satisfying \cref{def:ft-construction}, there is a range of physical error rates, i.e. \(\epsilon_{\text{P}} < \epsilon^*\), and fiducial rates \(\Delta\), which can be tolerated.
  In general, we may choose an \(\epsilon_{\text{P}}\) and \(\Delta\) so long as it is within the ranges allowable by the error correcting circuit.
  In practice, \(\epsilon_{\text{P}}\) is chosen by physical constraints (i.e. to minimize resource overhead), and \(\Delta\) is chosen to minimize the previously discussed width overhead.
  By \cref{def:ft-construction}iii, the error correcting circuit has constant a depth \(D\) independent of code size, and \(D\) is precisely the depth overhead.

  \emph{Dependency overhead---}
  Finally, since a circuit's outputs are in general statistically dependent, the width overhead may need to be scaled to appropriately counteract that dependence,
  We characterize the degree to which the error correcting circuits are independent by a factor \(\chi\), which we call the \emph{independency} (or equivalently the \emph{dependency} \(\chi^{-1}\)).
  In order for a family of error correcting circuits to be valid for a fault-tolerant construction, it must be that \(\chi > 0\).
  This is possible since, by \cref{def:ft-construction}, the family of error correcting circuits is comprised of gates of constant fan-in and constant depth which effectively limits the possible dependency of the circuit's outputs---of course this assume non-pathological wiring which we will discuss in \cref{ssec:ft-circuits}.

  This constant \(\chi\) is in general difficult to estimate as it depends on the specifics of the circuit construction, which becomes exponentially (with the number of wires in the circuit) to calculate exactly.
  In order to numerically estimate \(\chi\), we represent a noisy circuit as a Bayesian network where each wire is represented by a node, and a directed edge connects input wires to output wires of each gate;
  an example using the error correcting formula of \cite{evans1998on-the-maximum} and equivalent error correcting circuit of \cite{von-neumann1956probabilistic}, along with their induced Bayesian network representations is depicted in \cref{fig:bayesian-network}.
  Using this Bayesian network, we are able to estimate a circuit's performance to high precision and estimate its asymptotic behavior.

  In the following, we seek an estimate of the coefficient hidden in \cref{eq:d-asymptotics} (for small \(\epsilon_{\text{L}}\)), and an estimate of \(\chi\) for the fault-tolerant construction using \(2\)-input \nand~gates of \cref{rem:depth-2-ft-construction}.

\subsection{Formulas}
\label{ssec:ft-formulas}
  \begin{figure}
    \includegraphics[width=0.75\textwidth]{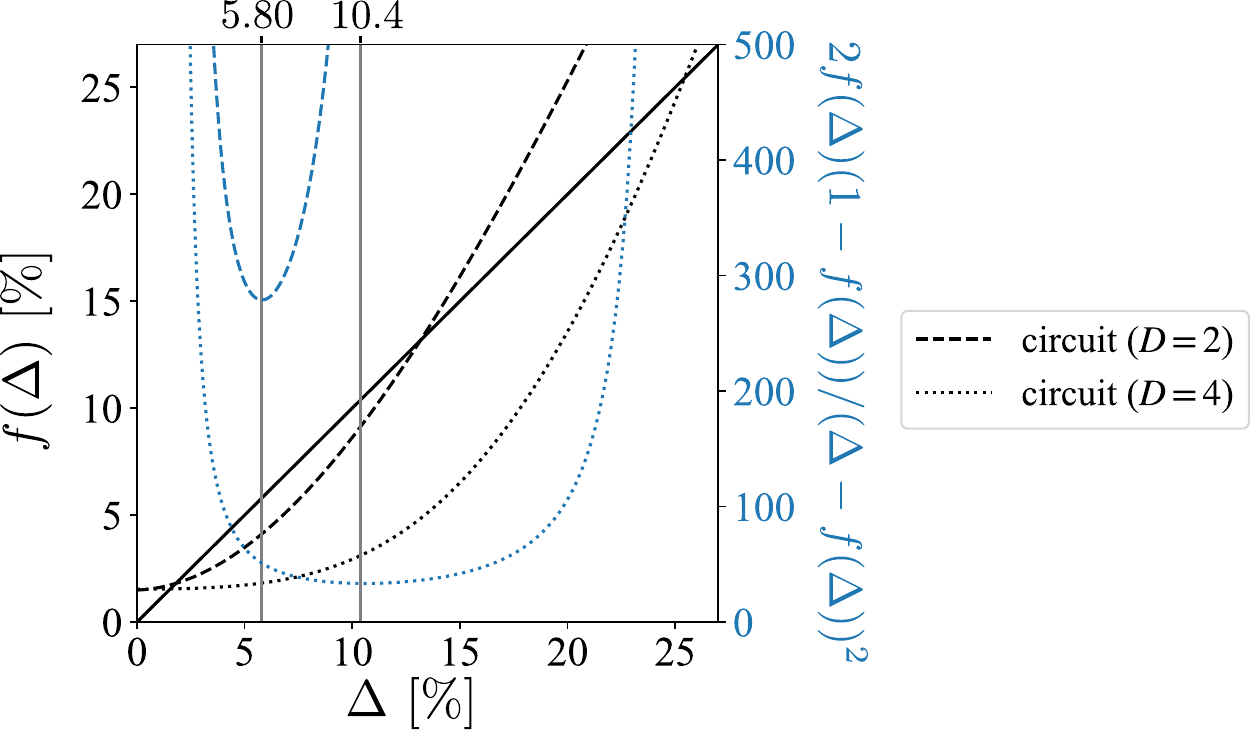}
    \caption{Output error rate $f_{\epsilon_P}(\Delta)$ (on the left $y$-axis) as a function input error rate $\Delta$  for \(\epsilon_{\text{P}} = 0.5\%\) from \cref{eq:formula-resultant-error-rate} compared to the \(y = \Delta\) line (solid black) for different error correction circuit constructions. The leading term in the code size $n(\epsilon_L; \epsilon_P, \Delta)$, the coefficient of \(\log(1 / \epsilon_{\text{L}})\) in \cref{eq:formula-asymptotic-code-size}, is shown on the right-hand $y$ axis and plotted with blue lines, using the same \(x\)-axis.  Its minima for the depth-\(2\) and depth-\(4\) majority formulas are marked (solid grey vertical lines). 
    For the depth-\(2\) majority formula (\cref{rem:depth-2-ft-construction}) this gives a minimum required code size of \(n \approx 279 \log(1 / \epsilon_{\text{L}})\), and for the depth-\(4\) majority formula (\cref{rem:larger-error-correcting-formulas}) this gives a minimum required code size of \(n \approx 11.4 \log(1 / \epsilon_{\text{L}})\).
    }
  \label{fig:computation-error-rate}
  \end{figure}

  We start with the simpler case of formulas for which outputs are independent. 
  In the case of formulas, the Bayesian network takes the form of a directed tree and each node is required to have out-degree \(1\) (except for the output(s) with out-degree \(0\)), the roots of all disjoint sub-trees are independent conditioned on their ancestors.
  Thus the analysis is simplified by the independence of the output wires.

  Following von Neumann's construction \cite{von-neumann1956probabilistic}, given \(4n\) independent copies of each signal \(X\) and \(Y\), one performs the \nand~gate \(4n\) times and takes the majority using the configuration of \cref{fig:bayesian-network}.
  Assuming each input wire errs with probability at most \(\Delta\), the output wire errs with probability bounded by
  \begin{equation}
  \label{eq:formula-resultant-error-rate}
    f_{\epsilon_{\text{P}}}(\Delta) = 1 - \epsilon_{\text{P}} + (2 \epsilon_{\text{P}} -1) \left((2 \epsilon_{\text{P}} -1) ((\Delta -2) \Delta  (2 \epsilon_{\text{P}} -1)+\epsilon_{\text{P}} )^2-\epsilon_{\text{P}} +1\right)^2,
  \end{equation}
  For large \(n\), we can approximate the binomial distributed output with a normal distribution as a result of the central limit theorem.
  Given the independence of the output wires, we have
  \begin{equation}
  \label{eq:normal-approximation-error-rate}
    \epsilon_{\text{L}}(n; \epsilon_{\text{P}}, \Delta) = \Pr\bracks*{Z \ge \sqrt{n} \parens*{\frac{\Delta - f_{\epsilon_{\text{P}}}(\Delta)}{\sqrt{f_{\epsilon_{\text{P}}}(\Delta) (1 - f_{\epsilon_{\text{P}}}(\Delta))}}}},
  \end{equation}
  where \(Z\) is a standard normal random variable, \(\Delta \in (0, 1/2)\) is a fiducial error rate, and we have dropped the subscript \(\epsilon_{\text{P}}\) for convenience.
  Using \cref{eq:normal-approximation-error-rate}, we find in the limit of large \(n\), 
  \begin{equation}
  \label{eq:formula-asymptotic-code-size-error-rate}
    \epsilon_{\text{L}} = \bracks*{\frac{f_{\epsilon_{\text{P}}}(\Delta) (1 - f_{\epsilon_{\text{P}}}(\Delta))}{\Delta - f_{\epsilon_{\text{P}}}(\Delta)} \frac{1}{\sqrt{2 \pi n}}+O\parens*{n^{-\frac{3}{2}}}} \exp\parens*{-\frac{n (\Delta - f_{\epsilon_{\text{P}}}(\Delta))^2}{2 f_{\epsilon_{\text{P}}}(\Delta) (1 - f_{\epsilon_{\text{P}}}(\Delta))}}.
  \end{equation}
  giving a code size,
  \begin{equation}
  \label{eq:formula-asymptotic-code-size}
    n(\epsilon_{\text{L}}; \epsilon_{\text{P}}, \Delta) = \frac{2 f_{\epsilon_{\text{P}}}(\Delta) (1 - f_{\epsilon_{\text{P}}}(\Delta))}{(f_{\epsilon_{\text{P}}}(\Delta) - \Delta)^2} \log\parens*{\frac{1}{\epsilon_{\text{L}}}} + O\parens*{\log\log\parens*{\frac{1}{\epsilon_{\text{L}}}}}.
  \end{equation}

  In practice, for a given architecture and \(\epsilon_{\text{P}}\), one chooses \(\Delta\) in order to maximize the coefficient of \(\sqrt{n}\) in \cref{eq:normal-approximation-error-rate}.
  For this fault-tolerant construction with output error rate bounded by \cref{eq:formula-resultant-error-rate}, we find a pseudothreshold of \(\epsilon^* \approx 1.077\%\).
  Given a physical error rate below this threshold, we may choose an operating point between the fixed-points of \cref{eq:formula-resultant-error-rate}.
  As an example, for \(\epsilon_{\text{P}} \approx 0.5\%\), these fixed points occur around \(\Delta \approx 1.81\%\) and \(\Delta \approx 13.2\%\);
  and we find from \cref{eq:normal-approximation-error-rate} an optimal fiducial rate of \(\Delta \approx 5.80\%\) as shown in \cref{fig:computation-error-rate} which implies a required code size of \(n \approx 279 \log(1 / \epsilon_{\text{L}})\).
  Further note that in this fault-tolerant construction, two of every three layers are dedicated to error correction, i.e. the depth overhead is \(D = 2\).

  \begin{remark}[Larger error correcting formulas]
  \label{rem:larger-error-correcting-formulas}
    In general, the pseudothreshold may be improved by increasing the size of the error correcting circuit.

    For the example of computing with \(2\)-input \nand~gates, the majority operation can be constructed using full binary trees of even depth \(D \ge 2\);
    for code size \(n\), this formula representation requires \(n(2^D - 1)\) gates.
    The depth-\(4\) majority formula has an increased pseudothreshold of \(\epsilon^* \approx 2.515\%\), strictly greater than that of the depth-\(2\) majority gate.
    In addition to the increased threshold, the larger majority formula also has different fixed points and a different optimal fiducial rate.
    For example, for \(\epsilon_{\text{P}} = 0.5\%\), the depth-\(4\) formula reduces input errors on expectation for fiducials between \(\Delta \approx 1.55\%\) and \(\Delta \approx 25.4\%\), with optimal performance for \(\Delta \approx 10.4\%\).
  \end{remark}

  While the analysis of formulas is easier and provides insight into the fault-tolerant construction, it is not suitable for implementation as the number of required gates grows exponentially with depth: 
  since sub-formulas must be duplicated each time its result is required in order to maintain the fan-out \(1\) condition.
  More realistically, by allowing larger fan-out, we can `fold' this formula representation into a more general circuit.

\subsection{Circuits}
\label{ssec:ft-circuits}
  \begin{figure}
    \includegraphics[width=0.45\textwidth]{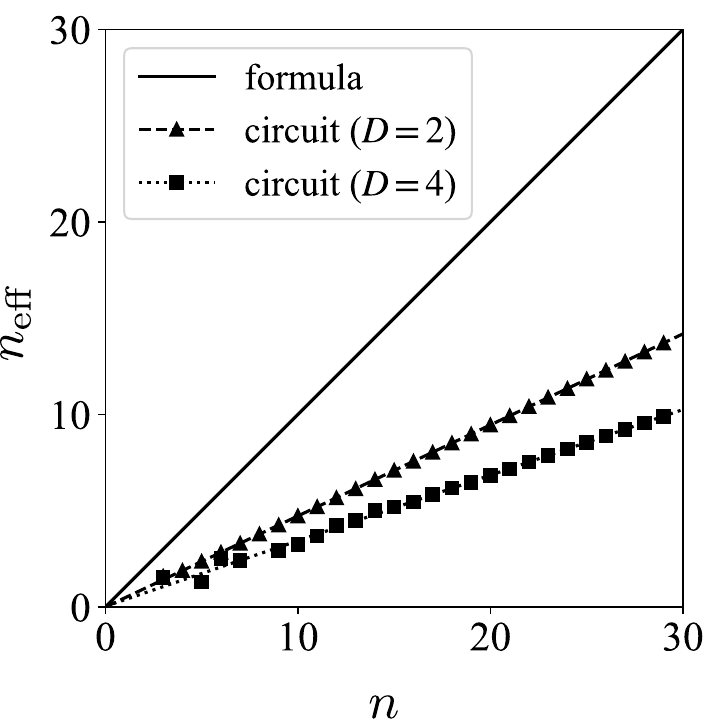}
    \caption{The effective code size for formulas and circuits for the fault-tolerant construction of \cref{rem:depth-2-ft-construction} and \cref{rem:larger-error-correcting-circuits} operating subject to physical error rate \(\epsilon_{\text{P}} = 0.5\%\) and at their optimal fiducial point.
    The slope of the formula (solid) line is \(1\) and the fitted lines for the \(D=2\) (dashed) and \(D = 4\) circuits (dotted). 
    We find \(n_{\text{eff}} \approx 0.47 n\) and \(n_{\text{eff}} \approx 0.34 n\) for the depth-\(2\) and depth-\(4\) error correction circuits respectively.
    }
  \label{fig:effective-size-comparison}
  \end{figure}

  \begin{figure}
    \includegraphics[width=0.65\textwidth]{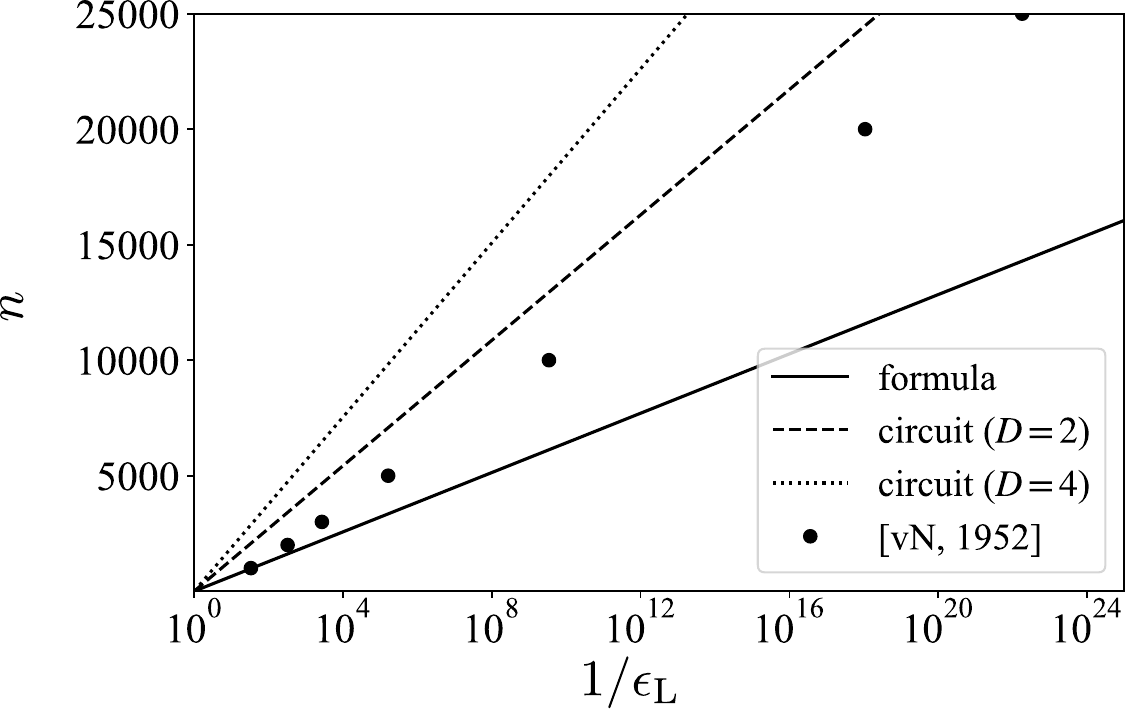}
    \caption{The code size required to achieve a target logical error rate \(\epsilon_{\text{L}}\) for odd \(n\) at \(\epsilon_{\text{P}} = 0.5\%\) for different majority circuit constructions at their respective optimal fiducial point.
    The code size required using the majority formula (solid) is \(n \approx 642 \log_{10}(1 / \epsilon_{\text{L}})\), for the depth-\(2\) majority circuit (dashed) code size is \(n \approx 1360 \log_{10}(1 / \epsilon_{\text{L}})\), and for the depth-\(4\) majority circuit (dotted) the code size is \(n \approx 1880 \log_{10}(1 / \epsilon_{\text{L}})\).
    Also plotted is von Neumann's approximation for a similar formula-based error correction scheme with the same physical error rate and slightly different choice of fiducial \(\Delta = 7.0\%\) \cite[Section 10.5.2]{von-neumann1956probabilistic}.
    }
  \label{fig:code-size-comparison}
  \end{figure}

  As previously discussed, the analysis for circuits is complicated by the fact that outputs are no longer necessarily independent.
  In addition to the width and depth overheads, the independency \(\chi \in (0, 1]\) contributes to the asymptotic number overhead of a fault-tolerant circuit.
  Intuitively, since output wires are no longer independent, logical error rates are higher than in equivalent formulas which by definition have \(\chi = 1\).

  From  \cref{def:ft-construction}iii, error correcting circuits are of constant depth, and therefore output wires are dependent only on constant number of input wires as depicted in \cref{fig:bayesian-network} (this is our main reason for including requirement iii, and excluding concatenation-based constructions).
  Therefore the same asymptotic scaling \(n = \Theta(\log(1 / \epsilon_{\text{L}}))\) holds as in the case with formulas but with a smaller effective code size \(n_{\text{eff}} \equiv \chi n\), where we call the constant \(\chi \in (0, 1]\) (\cref{def:ft-construction}vi precludes \(\chi = 0\)).
  In the case of the depth-\(2\) \nand~tree majority circuit, output wires are dependent on at most \(4\) input wires and therefore \(\chi \ge 1 / 4\).
  Use belief propagation to numerically approximate error rates using the induced Bayesian network representation (see \cref{fig:bayesian-network}), we find an independency of \(\chi \approx 0.47\) for the depth-\(2\) error correction circuit (see \cref{fig:effective-size-comparison}).
  In general \(\chi\) is dependent on the error correction circuit and choice of fiducial \(\Delta\).

  Of course, larger error correcting circuits, beyond the depth-\(2\) construction of \cref{rem:depth-2-ft-construction}, can also be used to perform error correction in a fault-tolerant construction.
  We provide a discussion of such circuits below:
  \begin{remark}[Larger error correcting circuits]
  \label{rem:larger-error-correcting-circuits}
    The majority formulas of \cref{rem:larger-error-correcting-formulas} can be `wrapped' into depth-\(D\) majority circuits using \(nD\) \nand~gates.
    For optimal performance, a path should connect each output gate to the maximal number, i.e. \(2^D\), of distinct input wires (for simplicity assume \(n \ge 2^D\)).
    If we label each gate by a tuple \((\ell, i)\) for layer \(\ell \in \{1, \dots, D\}\) and index \(i \in \{0, \dots, n-1\}\), one such construction is to connect gate \((\ell, i)\) to the output of gates \((\ell - 1, i)\) and \((\ell - 1, i - 2^{\ell-1} \mod{n})\);
    where \((0, i)\) denotes the \(i\textsuperscript{th}\) input wire.

    As with the depth-\(2\) case (\cref{rem:depth-2-ft-construction}), larger majority circuits are less effective at error correction than their formula counterparts due to the dependence of the output signals.
    Numerics show an independency of \(\xi \approx 0.34\) for depth-\(4\) circuits operating at \(\epsilon_{\text{P}} = 0.5\%\) at the optimal fiducial rate.

    It is widely conjectured, though not rigorously proven, that the pseudothreshold of such a circuit approaches the Evans--Pippenger rate of \(\epsilon_{\text{EP}} = (3 - \sqrt{7}) / 4 \approx 0.08856\) as we take \(D \rightarrow \infty\) \cite{von-neumann1956probabilistic,evans1998on-the-maximum,unger2007noise}.

    Note that there are other ways to fold the formulas into circuits which may result in poor performance.
    For example, imagine that all outputs were made dependent on the same \(2^D\) inputs independently of \(n\).
    In this case, we will find that increasing the code size does not decrease the error and \(\chi = 0\).
    This is true despite the fact that it shares the same unfolded formula representation as the construction described above.
    Though this is a pathological example, one may find that some reasonable circuit constructions result in lower independency.
  \end{remark}

  Notice that each computation gate in the non-fault-tolerant circuit is repeated \(n(\epsilon_{\text{L}}; \epsilon_{\text{P}}, \Delta)\) times in the fault-tolerant circuit, and followed by an error correction circuit of size \(D \times n(\epsilon_{\text{L}}; \epsilon_{\text{P}}, \Delta)\).
  Therefore, for fixed \(\epsilon_{\text{P}}\), the gate overhead is
  \begin{equation}
  \label{eq:number-overhead}
    \eta_{\#}(\epsilon_{\text{L}}; \epsilon_{\text{P}}, \Delta, D, \chi) = \frac{1}{\chi} (D + 1) n(\epsilon_{\text{L}}; \epsilon_{\text{P}}, \Delta),
  \end{equation}
  where we use \(\eta_{\#}\) to denote the overhead in the number of gates.
  When a particular error correction construction is assumed, we may drop explicit dependence on the variables to the right of the semicolon in the arguments of \(\eta_{\#}\) and \(n\), i.e. \(D\), \(\chi\), \(\epsilon_{\text{P}}\), and \(\Delta\).

  To summarize, the asymptotic number overhead of a fault-tolerant circuit can be deconstructed into three contributions:
  \begin{itemize}[noitemsep,topsep=0pt]
    \item the width or code size overhead \(n(\epsilon_{\text{L}}; \epsilon_{\text{P}}, \Delta)\), which can be obtained through the analysis of its equivalent formula representation;
    \item the depth overhead, i.e. the constant in \cref{def:ft-construction}iii; and,
    \item the dependency overhead \(\chi^{-1}\), which is a function of how the formula is `folded' into a circuit, and can be obtained by numerical means (e.g. using belief propagation or Monte Carlo sampling).
  \end{itemize}

\section{Fault-tolerance resource overhead}
\label{sec:resource-overhead}
  Having derived the number overhead (\cref{eq:number-overhead}) of the fault-tolerant construction of \cref{rem:depth-2-ft-construction}, we turn to its resource overhead.
  To this end, we introduce the notion of a \emph{resource--reliability trade-off} which associates a resource cost for each gate as a function of the physical error rate.
  In order to formalize the problem, let \(W: (0, \epsilon^*) \rightarrow \R_{\ge 0}\) denote the resource--reliability trade-off: \(W(\epsilon_{\text{P}})\) represents the resource cost of a single gate operating with physical error rate \(\epsilon_{\text{P}}\).

  Using the resource--reliability trade-off in conjunction with the number overhead (\cref{sec:gate-overhead}), we study the cases in which either the fault-tolerant or non-fault-tolerant constructions may be preferred in order to reduce overall resource utilization.
  While the specific resources of interest are application-specific---e.g. power dissipation \cite{impens2004fine-grained,thaker2005recursive,chatterjee2016energy-reliability,chatterjee2020energy-reliability}, layout area \cite{impens2004fine-grained,thaker2005recursive,gadlage2006digital,oates2015reliability} for transistor-based gates, or even economic costs---we find general conclusions that depend only on the asymptotic behavior of this resource--reliability trade-off in the low-error limit \(\epsilon \rightarrow 0\).

  We make the physical assumption that \(W(\epsilon_{\text{P}})\) is monotonically non-increasing with respect to \(\epsilon_{\text{P}}\): that is, a reduced physical error rate necessitates equal or larger resource expenditure.
  With the resource--reliability function in hand, we can state our main asymptotic result:
  \begin{remark}[Fault-tolerant overhead theorem]
  \label{rem:asymptotic-fault-tolerance-overhead}
    Suppose we have a fault-tolerant construction with number overhead \(\eta_\#(\epsilon_{\text{L}}; \epsilon_{\text{P}}, \Delta, D, \chi)\), where \(\epsilon_{\text{L}}\) is the target logical error rate and \(\epsilon_{\text{P}}\), \(\Delta\), \(D\), and \(\chi\) are parameters of the fault-tolerant construction.
    In the limit of small logical error rates \(\epsilon_{\text{L}} \rightarrow 0\), the non-fault-tolerant construction is preferred for resource--reliability trade-offs which are asymptotically \(W(\epsilon_{\text{P}}) = o(\log(1 / \epsilon_{\text{P}}))\) as \(\epsilon_{\text{P}} \rightarrow 0\).
    Conversely, in the same small logical error rate limit, the fault-tolerant construction is preferred for resource--reliability trade-offs which are asymptotically \(W(\epsilon_{\text{P}}) = \omega(\log(1 / \epsilon_{\text{P}}))\) as \(\epsilon_{\text{P}} \rightarrow 0\).
  \end{remark}
  \cref{rem:asymptotic-fault-tolerance-overhead} is a statement about the low logical error rate asymptotics of the resource--reliability trade-off; however, for most practical purposes it is sufficient to achieve a low, but finite, logical error rate.
  For a sense of scale, a study estimated error rates of \(\sim 10\) errors per billion hours of operation per gate in modern transistor-based logic \cite{wang2008soft}.
  Assuming such gates perform \(10^9\) operations per second, this corresponds to a logical per gate error rate of \(\epsilon_{\text{L}} \approx 3 \times 10^{-21}\).
  In the subsequent analysis, we estimate the benefit of fault-tolerance (or lack thereof) assuming values for the behavior of the resource--reliability trade-off required to achieve logical error rates down to this level.

  To further motivate the resource--reliability trade-off, \(W(\epsilon_{\text{P}})\), consider that the output of each gate is a bit \(x \in \{0, 1\}\) encoded in some real-valued physical degree of freedom \(\{-R, +R\}\), where \(R > 0\) denotes the signal strength.
  Suppose the signal is corrupted by noise \(e\), which for our purposes will be a symmetric, zero mean random variable, so that our overall signal is \(X_\pm = \pm R + e\) where \(e\) denotes an error term.
  We consider \(X_\pm\) to be correct if it lies on the correct side of \(0\), and therefore the physical error rate is \(\epsilon_{\text{P}}(R) = \Pr[X_+ < 0 | R] = \Pr[X_- > 0 | R]\).
  Further assume that the noise \(e\) is i.i.d. at the output of each gate, and that the resource utilization is proportional to \(W \propto L\).

  This model allows us to relate the resource--reliability trade-off \(W(\epsilon_{\text{P}})\) to the distribution of the noise \(e\).
  For our setting of interest, the resource utilization in the limit of low logical error rates, there are broadly three scenarios to consider: exponentially-tailed, light-tailed, and heavy-tailed noise distribution. 
  We discuss these cases in sequence below.

  \subsection{Exponentially-tailed noise distribution}
  \label{ssec:exp-tailed}
  \begin{figure}
    \includegraphics[width=0.6\textwidth]{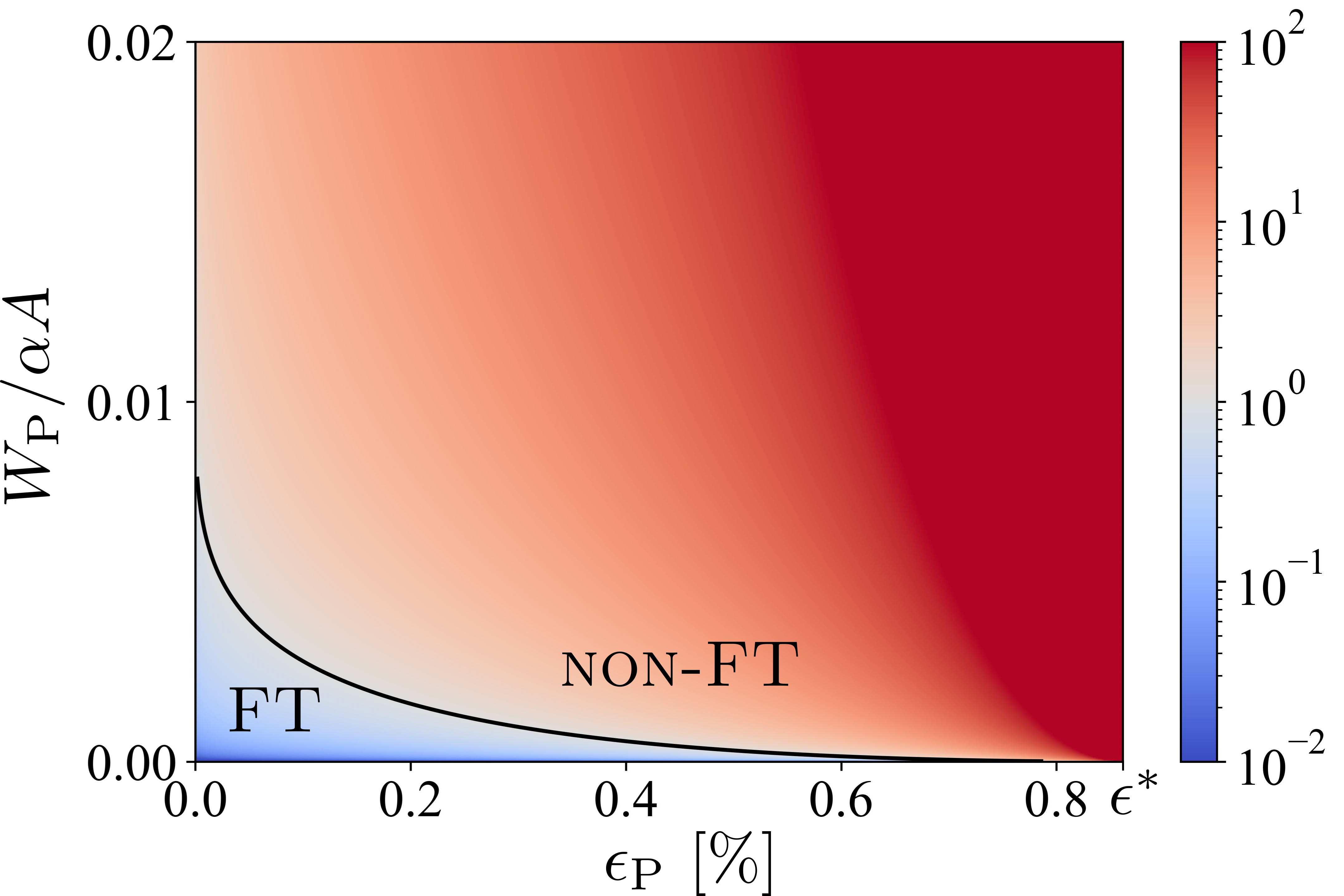}
    \caption{Plot of the asymptotic fault-tolerance overhead in the \(\epsilon_{\text{L}} \rightarrow 0\) limit for exponentially-tailed resource functions as predicted by \cref{eq:fault-tolerance-overhead-ratio-exponential-tail} for a fault-tolerant construction based on the depth-\(2\) majority circuit of \cref{rem:depth-2-ft-construction} and optimal fiducial \(\Delta\).
    The region where the fault-tolerant construction is less resource intensive is marked (\textsc{FT}) and its complement is marked (\textsc{non-FT}).  
    }
  \label{fig:phase-diagram-exponential-tail}
  \end{figure}

  First, consider the marginal case where the resource--reliability trade-off has the same asymptotic scaling as the number overhead, i.e. \(W(\epsilon_{\text{P}}) = \Theta(\log(1 / \epsilon_{\text{P}}))\).
  In our abstract model, this would correspond to an error \(e\) with exponentially distributed tails so that as \(\epsilon_{\text{P}} \rightarrow 0\) or equivalently \(R \gg \alpha\), we have
  \begin{equation}
  \label{eq:exponential-tail}
    \epsilon_{\text{P}} \sim C \exp\parens*{-\frac{R}{\alpha}},
  \end{equation}
  for some constants \(\alpha > 0\) and \(C > 0\).

  In this case as \(\epsilon \rightarrow 0\), 
  \begin{equation}
  \label{eq:exponential-tail-resource-function}
    W(\epsilon_{\text{P}}) \sim \alpha A \log\parens*{\frac{1}{\epsilon_{\text{P}}}},
  \end{equation}
  where \(A > 0\) is a proportionality constant.

  Combining the number overhead (\cref{eq:number-overhead}) and assuming exponentially-tailed resource--reliability trade-off (\cref{eq:exponential-tail-resource-function}), we find the resource overhead as \(\epsilon_{\text{L}} \rightarrow 0\) required for fault-tolerance to be
  \begin{align}
  \label{eq:fault-tolerance-overhead-ratio-exponential-tail}
    \eta(\epsilon_{\text{L}}; \epsilon_{\text{P}}, \Delta, D, \chi) &= \frac{W(\epsilon_{\text{P}})}{W(\epsilon_{\text{L}})} (D + 1) n(\epsilon_{\text{L}}; \epsilon_{\text{P}}, \Delta) \\
     &= \parens*{\frac{W_{\text{P}}}{\alpha A}} \parens*{\frac{2 (D + 1) f_{\epsilon_{\text{P}}}(\Delta) (1 - f_{\epsilon_{\text{P}}}(\Delta))}{\chi (f_{\epsilon_{\text{P}}}(\Delta) - \Delta)^2}},
  \end{align}
  where \(W_{\text{P}} > 0\) represents the resource utilization at finite physical error rate \(\epsilon\).
  As with \cref{eq:number-overhead}, when a particular error correction construction is assumed, we may drop explicit dependence on the variables to the right of the semicolon in the arguments of \(\eta\).

  Note the fault-tolerance overhead as \(\epsilon_{\text{L}} \rightarrow 0\) is independent of \(\epsilon_{\text{L}}\) in this case of an exponentially-tailed error function (or resource--reliability trade-off satisfying \cref{eq:exponential-tail-resource-function}).
  In the case of exponentially-tailed errors, both increasing \(R\) and increasing the code size \(n\) (assuming a linear distance code) suppress errors exponentially; and thus, which approach is preferred to achieve a logical error rate \(\epsilon_{\text{L}} \rightarrow 0\) depends on the specifics of both the resource--reliability trade-off and the fault-tolerant construction.
  In \cref{eq:fault-tolerance-overhead-ratio-exponential-tail} relevant terms have been grouped into those that depend on the asymptotic behavior of the resource utilization model: \(W_{\text{P}}\), \(\alpha\) and \(A\); 
  and those that depend on the fault-tolerant circuit construction: \(\epsilon_{\text{P}}\), \(f_{\epsilon_{\text{P}}}\), \(\Delta\), \(D\), and \(\chi\).
  The overhead is plotted in \cref{fig:phase-diagram-exponential-tail}. Regions are indicated where one construction, i.e. fault-tolerant or non-fault-tolerant, is preferred.

  \subsection{Light-tailed resource function}
  \label{ssec:light-tailed}
  \begin{figure}
    \includegraphics[width=0.6\textwidth]{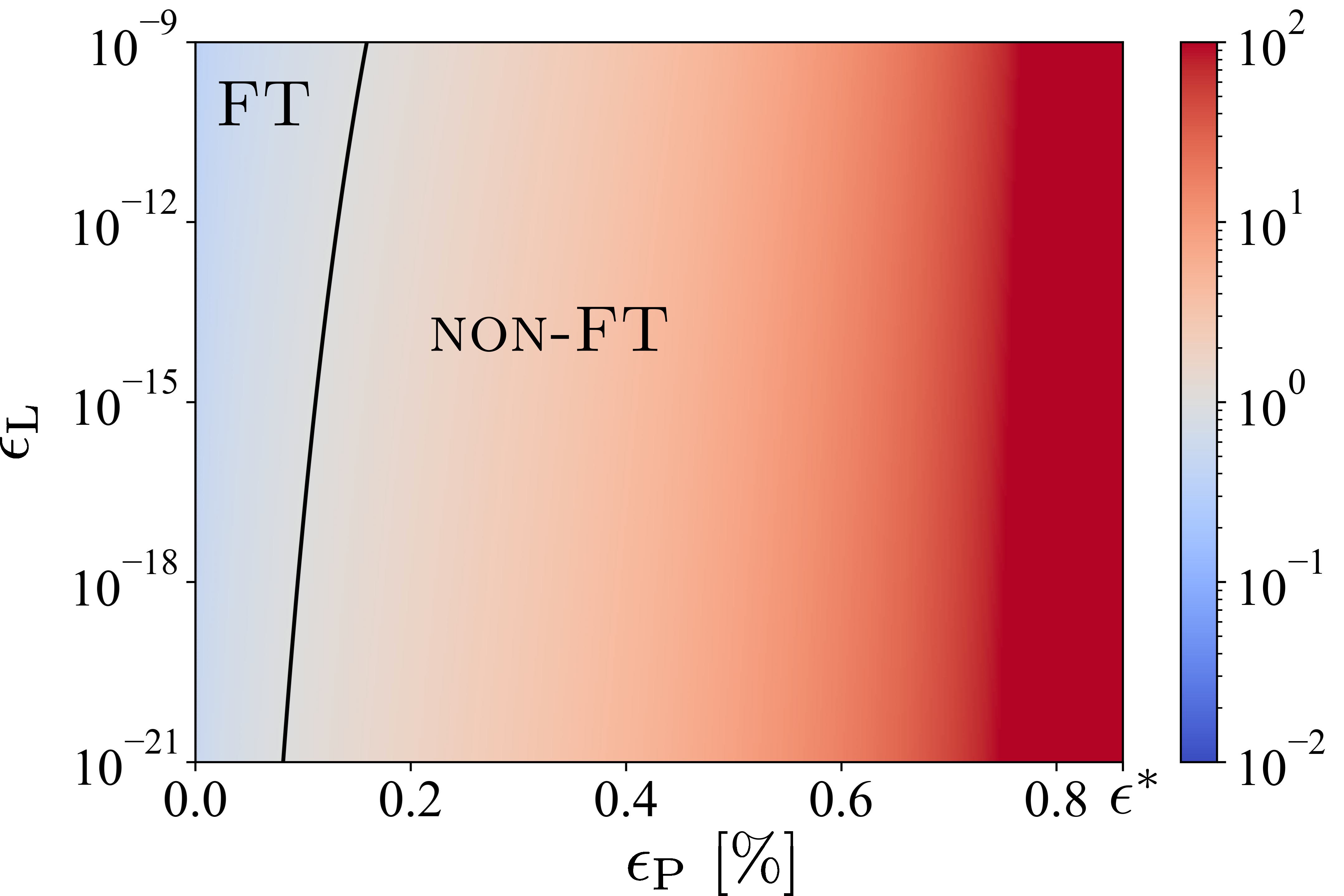}
    \caption{Plot of the asymptotic fault-tolerance overhead in the \(\epsilon_{\text{L}} \rightarrow 0\) limit for Gaussian-tailed errors as predicted by \cref{eq:fault-tolerance-overhead-ratio-light-tail} for a fault-tolerant construction based on the depth-\(2\) majority circuit of \cref{rem:depth-2-ft-construction}, optimal fiducial \(\Delta\), and with constants \(W_{\text{P}} / \sqrt{2} \sigma A = 2 \times 10^{-4}\).
    The region where \(\eta < 1\) indicates the regions where the fault-tolerant construction is less resource intensive is marked (\textsc{FT}) and its complement is marked (\textsc{non-FT}).
    It is possible that no \textsc{FT} region exists for some settings of the constants \(W_{\text{P}} / \sqrt{2} \sigma A\).
    }
  \label{fig:phase-diagram-light-tail}
  \end{figure}

    In the case of light-tailed error distributions, i.e. those whose tails are super-exponential, we will show that the non-fault-tolerant construction is always preferred as the target error rate \(\epsilon_{\text{L}} \rightarrow 0\).
    For simplicity, consider the case of an error distribution with a Gaussian tail.
    For \(R \ge 0\),
    \begin{equation}
    \label{eq:gaussian-tail}
      \epsilon_{\text{P}}(R) 
      \sim \bracks*{\frac{C}{\sqrt{2 \pi} R} + O\parens*{\frac{1}{R^3}}} \exp\parens*{-\frac{R^2}{2 \sigma^2}},
    \end{equation}
    for constants \(C > 0\) and \(\sigma > 0\) as \(R \rightarrow \infty\).

  In this case as \(\epsilon \rightarrow 0\), 
  \begin{equation}
  \label{eq:gaussian-tail-resource-function}
    W(\epsilon_{\text{P}}) \sim \sqrt{2} \sigma A \operatorname{erfc}^{-1}(2 \epsilon_{\text{P}}),
  \end{equation}
  where \(A > 0\) is a proportionality constant.

  Using the number overhead (\cref{eq:number-overhead}), we find the following expression for the overhead in the case of a Gaussian-distributed error \(e\),
  \begin{align}
  \label{eq:fault-tolerance-overhead-ratio-light-tail}
    \eta_{\text{Gaussian}}(\epsilon_{\text{L}}; \epsilon_{\text{P}}, \Delta, D, \chi)
     &= 
     \parens*{\frac{W_{\text{P}}}{\sqrt{2} \sigma A \operatorname{erfc}^{-1}(2 \epsilon_{\text{L}})}}
     \parens*{\frac{2 (D + 1) f_{\epsilon_{\text{P}}}(\Delta) (1 - f_{\epsilon_{\text{P}}}(\Delta))}{\chi (f_{\epsilon_{\text{P}}}(\Delta) - \Delta)^2} \log(1 / \epsilon_{\text{L}})},
  \end{align}
  where \(W_{\text{P}}\) is the resource utilization at finite physical error rate \(\epsilon_{\text{P}}\).

  From \cref{eq:fault-tolerance-overhead-ratio-light-tail} we observe that in the \(\epsilon_{\text{L}} \rightarrow 0\) limit, \(\eta_{\text{Gaussian}} > 1\), and therefore the non-fault-tolerant construction is preferred.
  It is nonetheless possible that for some finite \(\epsilon_{\text{L}}\), there exists a setting of \(\epsilon_{\text{P}}\) where the fault-tolerant construction is more resource efficient in this case as the efficiency at finite logical error rates depends on \(W_{\text{P}} / \sqrt{2} \sigma A\), which depends on resource utilization at finite physical error rates.
  One setting of parameters that admits a region of fault-tolerant resource savings is shown in \cref{fig:phase-diagram-light-tail}.

  Light-tailed error distributions, can arise in a number of scenarios.
  Gaussian-distributed noise may occur naturally as a result of the central limit theorem when noise sources are the sum of many independent components.
  The fundamental physical limits on power dissipation are themselves light-tailed resource--reliability trade-offs.
  Landauer famously argued that in an idealized computer, the only dissipation necessary is proportional to the amount of logical irreversibility \cite{landauer1961irreversibility};
  Fredkin and Toffoli subsequently showed that computations may be designed to require only constant amounts of logical irreversibility \cite{fredkin1982conservative}.
  Given a well isolated system with idealized control of microscopic degrees of freedom, one would find a resource--reliability trade-off such that \(\lim_{\epsilon \rightarrow 0} W(\epsilon) = C\) for some constant \(C\) (the primary resource of concern in this case would likely not be power dissipation but rather the economic cost of engineering such a system).

  We note that while these results are for Gaussian-tailed noise, the qualitative result---namely that the non-fault-tolerant construction is preferred in the \(\epsilon_{\text{L}} \rightarrow 0\) limit---hold as long as the noise has super-exponential tails.
  This can be understood as physical error rates may be suppressed through by increased resource utilization faster than the exponential suppression of logical error rates using a linear-distance error correcting code.

  \subsection{Heavy-tailed resource function}
  \label{ssec:heavy-tailed}
  \begin{figure}
    \includegraphics[width=0.6\textwidth]{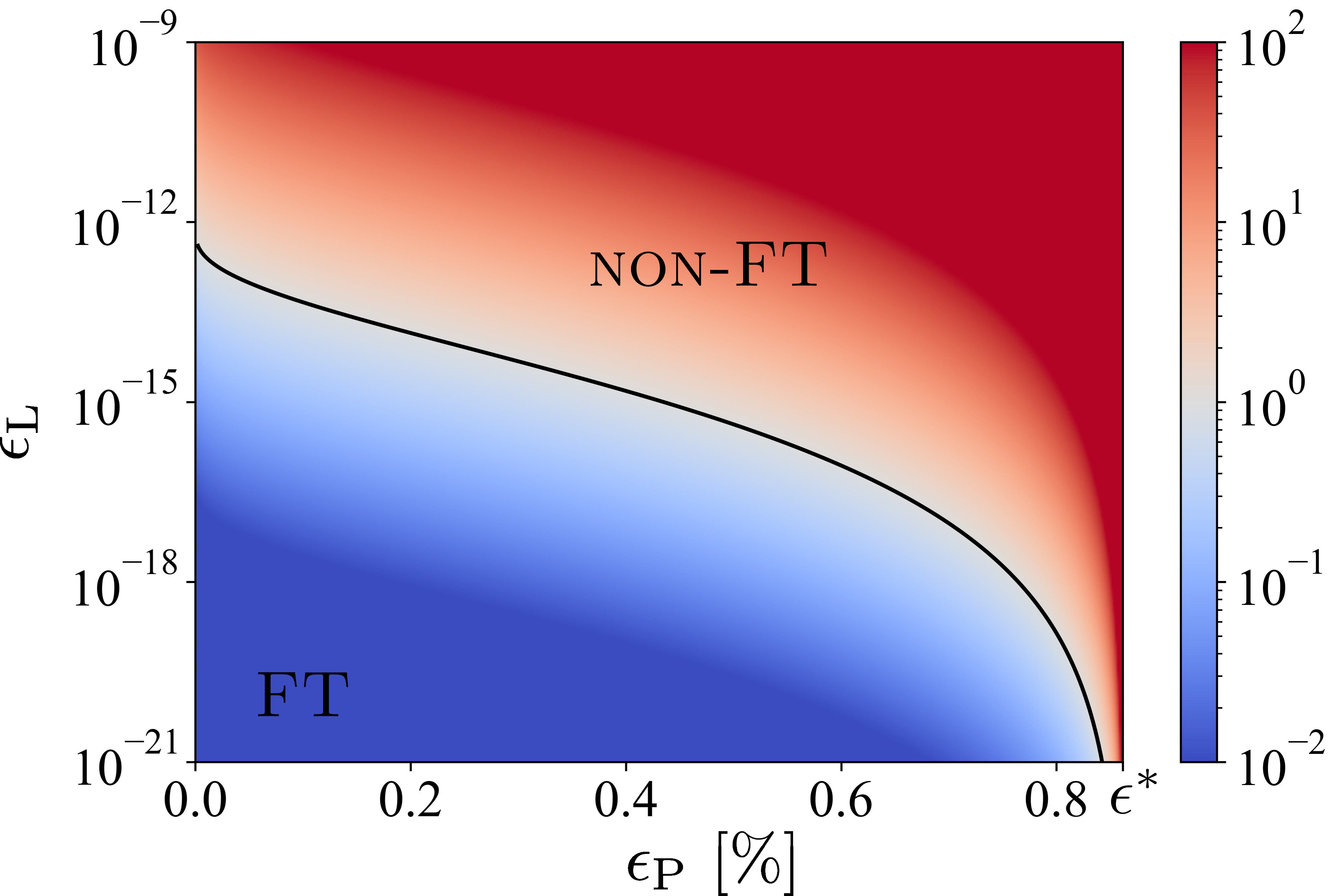}
    \caption{Plot of the asymptotic fault-tolerance overhead in the \(\epsilon_{\text{L}} \rightarrow 0\) limit for Pareto-tailed errors as predicted by \cref{eq:fault-tolerance-overhead-ratio-heavy-tail} for a fault-tolerant construction based on the depth-\(2\) majority circuit of \cref{rem:depth-2-ft-construction}, optimal fiducial \(\Delta\), \(\gamma = 2\), and constants \(W_{\text{P}} / A \beta = 3 \times 10^2\).
    The region where the fault-tolerant construction is less resource intensive is marked (\textsc{FT}) and its complement is marked (\textsc{non-FT}). 
    In this case of a heavy-tailed resource function, the fault-tolerant construction is always preferred as the target error rate goes to zero.
    }
  \label{fig:phase-diagram-heavy-tail}
  \end{figure}

    In the case of heavy-tailed error distributions, i.e. those whose tails are sub-exponential, we find that the fault-tolerant construction is always preferred as the target error rate \(\epsilon_{\text{L}} \rightarrow 0\) (\cref{rem:asymptotic-fault-tolerance-overhead}).
    For simplicity, assume that the error \(e\) is distributed according to a symmetric Pareto distribution so that for \(R \ge 0\),
    \begin{equation}
    \label{eq:pareto-tail}
      \epsilon_{\text{P}}(R) = \frac{1}{2} \parens*{\frac{\beta}{\beta + R}}^\gamma,
    \end{equation}
    for some constants \(\beta > 0\) and \(\gamma > 0\).

  In this case for \(\epsilon_{\text{P}} \rightarrow 0\) we have,
  \begin{equation}
  \label{eq:pareto-tail-resource-function}
    W(\epsilon_{\text{P}}) \sim A \beta \parens*{\frac{1}{(2 \epsilon_{\text{P}})^{1 / \gamma}} - 1},
  \end{equation}
  where \(A > 0\) is again a proportionality constant.

  Using the number overhead (\cref{eq:number-overhead}), we find the following expression for the overhead in the case of a Pareto-tailed error \(e\),
  \begin{align}
  \label{eq:fault-tolerance-overhead-ratio-heavy-tail}
    \eta(\epsilon_{\text{L}}; \epsilon_{\text{P}}, \Delta, D, \chi)_{\text{Pareto}, \gamma}
     &= 
     \parens*{\frac{W_{\text{P}}}{A \beta} \frac{(2 \epsilon_{\text{L}})^{1 / \gamma}}{1 - (2 \epsilon_{\text{L}})^{1 / \gamma}}}
     \parens*{\frac{2 (D + 1) f_{\epsilon_{\text{P}}}(\Delta) (1 - f_{\epsilon_{\text{P}}}(\Delta))}{\chi (f_{\epsilon_{\text{P}}}(\Delta) - \Delta)^2}  \log(1 / \epsilon_{\text{L}})},
  \end{align}
  where \(W_{\text{P}}\) is again the resource utilization at finite physical error rate \(\epsilon_{\text{P}}\).

  From \cref{eq:fault-tolerance-overhead-ratio-heavy-tail} we observe that in the \(\epsilon_{\text{L}} \rightarrow 0\) limit, \(\eta_{\text{Pareto}} < 1\), and therefore the fault-tolerant construction is preferred.
  A representative plot of \(\eta_{\text{Pareto}}\) is shown in \cref{fig:phase-diagram-heavy-tail}.

  There are a number of ways in which a heavy-tailed error distribution may arise.
  One example discussed by Thaker et al. is the case where the resource of interest is of CMOS feature sizes, where the salient resource is the feature area \cite{thaker2005recursive}.
  Cohen et al. demonstrated experimentally that CMOS technology exposed to high levels of alpha radiation resulted in a resource--reliability trade-off with asymptotically Pareto tails with \(\gamma \approx 2.5\) \cite{cohen1999soft}.

  We note that while these results are for symmetric Pareto-distributed noise, the qualitative result---namely that the fault-tolerant construction is preferred in the \(\epsilon_{\text{L}} \rightarrow 0\) limit---hold as long as the noise has sub-exponential tails.
  This can be understood as (sufficiently) independent repetitions of a calculation in this case suppress logical error rates exponentially, faster than is achievable by increasing the resource utilization of individual gates under this resource--reliability trade-off.

\section{Concluding Remarks}
\label{sec:concluding-remarks}
  We have devised a method to perform an asymptotic analysis of the number of additional gates required by fault-tolerant constructions compared to the non-fault-tolerant circuits from which they are derived.
  Using this accounting, we guide the design of resource-efficient fault-tolerant circuits based on the resource--reliability trade-off of the basic unit.

  In \cref{sec:gate-overhead}, we first formalized the notion of a fault-tolerant construction.
  Key to our definition was the restriction to constant-depth error correcting circuits (\cref{def:ft-construction}iii), which is in contrast to the recursive concatenation-based approaches whose overheads have previously been studied.
  In concatenation-based constructions, the fault-tolerant circuit is constructed through recursive application of the \(\operatorname{FT}_n\) of \cref{def:ft-construction} at the level of the logical gate \cite{nielsen2010quantum,impens2004fine-grained,thaker2005recursive,svore2005a-flow-map} --- usually for fixed \(n\), e.g. \(n = 3\) in the recursive triple modular redundancy constructions of \cite{impens2004fine-grained,thaker2005recursive}.
  While the concatenation-based approach is useful for the analysis of asymptotic thresholds through intermediate pseudothreshold, e.g. using a flow-map model \cite{svore2005a-flow-map}, it conflates what we have termed the depth and width overheads since concatenation increases both simultaneously.
  For the analysis at fixed and finite \(\epsilon_{\text{P}}\) where we are satisfied to accept pseudothresholds (instead of true asymptotic thresholds), we may fix the depth \(D\) of the error correcting circuit while nonetheless increasing code size \(n\) thereby decreasing the logical error rate \(\epsilon_{\text{L}}\).
  This subtly in \cref{def:ft-construction} allowed us to interpret the number overhead as coming from three distinct components: the width overhead, the depth overhead, and the dependency overhead (\cref{ssec:scaling-arguments}).
  While the width and the depth overheads are most readily analyzed using fault-tolerant formulas (\cref{ssec:ft-formulas}), the dependency overhead captures the deviation from the formula resulting from the realities of the circuit model and was numerically estimated using inference techniques on the Bayesian networks induced by the noisy circuit (\cref{ssec:ft-circuits}).

  In \cref{sec:resource-overhead}, we formally introduced the resource--reliability trade-off and stated the main asymptotic result (\cref{rem:asymptotic-fault-tolerance-overhead}) which depends crucially on the \(\epsilon_{\text{P}} \rightarrow 0\) behavior of the resource--reliability curve.
  We studied the regions in the fault-tolerant construction parameter space in which fault-tolerance may be preferred based on assumptions about the finite-\(\epsilon_{\text{P}}\) behavior of the resource--reliability function.
  We show three qualitatively different results:
  in the case where \(W(\epsilon_{\text{P}}) = \Theta(\log(1 / \epsilon_{\text{P}}))\), the existence of a region of fault-tolerant resource savings is asymptotically independent of logical error rate \(\epsilon_{\text{L}}\) and depends solely on constant factors (\cref{ssec:exp-tailed});
  if \(W(\epsilon_{\text{P}}) = o(\log(1 / \epsilon_{\text{P}}))\), the existence of a region of fault-tolerant resource savings may be possible for relatively large values of \(\epsilon_{\text{L}}\) depending on finite-\(\epsilon_{\text{P}}\) behavior (\cref{ssec:light-tailed}); and 
  if \(W(\epsilon) = \omega(\log(1 / \epsilon))\), then the fault-tolerant construction always preferred for sufficiently small \(\epsilon_{\text{L}}\) (\cref{ssec:heavy-tailed}).

  \emph{Beyond the repetition-code---}
  The repetition code has been the subject of most inquiry into classical fault-tolerant constructions, with little attention given to the study of other codes \cite{von-neumann1956probabilistic,hajek1991on-the-maximum,pippenger1988reliable,evans2003on-the-maximum}.
  This is in stark contrast with the study of fault-tolerant models of quantum computation, where the development and analysis of new codes is a primary research thrust \cite{campbell2017roads}.
  The repetition code has largely been unchallenged in the classical setting in part due to matching negative results showing that the repetition code is sufficient to yield the optimal threshold error rate for commonly studied gate sets \cite{evans1998on-the-maximum,evans2003on-the-maximum}.
  Furthermore, the repetition code's simplicity yields both constant-depth (transversal) logical operations and constant-depth error correction, both of which are all the more important in the classical setting as no error-free computation is allowed; 
  in contrast, the typical setting for fault-tolerant quantum schemes assumes access to noiseless classical computation.
  However, the analysis of fault-tolerance overheads, particularly the constant factor analysis, may provide another way to compare fault-tolerant constructions in which other codes may be more favorable than the repetition code.

  Notably, our restriction to bounded depth error correction circuits also limits us to codes which can be corrected locally, i.e. only observing a constant number of input wires \cite{yekhanin2012locally}.
  This self-imposed restriction may be lifted by loosening \cref{def:ft-construction}iii to allow the error correcting circuit's depth to grow with the code size at the cost of introducing \(\epsilon_{\text{L}}\)-dependency to the depth overhead and affecting the asymptotic result of \cref{rem:asymptotic-fault-tolerance-overhead}.

  \emph{Fault-tolerance in other systems---}
  Returning to the motivation for von Neumann's original work \cite{von-neumann1956probabilistic}: biological systems perform noisy information processing and seems natural to ask whether they take advantage from some form of fault-tolerance.
  While the highly-structured von Neumann construction for fault-tolerance, i.e. alternation of computation and error correction, may be unrealistic in most systems, general principles may hold and may indeed be preferred due to more favorable resource efficiency or reliability.
  For example, error correction mechanisms have been discovered in biology \cite{forsdyke1981are-introns,sreenivasan2011grid,battail2019error-correcting}, and some studies have looked into fault-tolerance of biologically-inspired network models of computation \cite{zlokapa2022biological}.
  Signatures of error correction and fault-tolerance have also been observed to emerge in noisy Boolean networks made to undergo evolution-like dynamics \cite{mccourt2023noisy}.
  Given the ubiquity of error correction in biology, might it be possible that there are also signatures of fault-tolerance?
  Resource efficiency may be one way to probe the development of both error correction and fault-tolerance in biological systems.

  In addition to naturally occurring biological systems, there many engineered systems in which information processing occurs in a distributed and networked manner, e.g. electronic social networks and financial markets \cite{kirou2008computational}.
  Information flow in these networks are subject to noise and resource constraints.
  Furthermore, many of these systems, such as in financial markets, the noise is often heavy-tailed \cite{ibragimov2015heavy-tailed}, which is precisely the scenario where fault-tolerant design may be preferred.
  Again, the highly-structured von Neumann construction for fault-tolerance may be unrealistic in these cases, but it may be the case that principles of fault-tolerant design may provide resource savings in the design of complicated networks.

\begin{comment}
\begin{acknowledgments}
  The work of AKT and ILC on analysis and numerical simulation was supported by the U.S. Department of Energy, Office of Science, National Quantum Information Science Research Centers, Co-design Center for Quantum Advantage (C$^2$QA) under contract number DE-SC0012704.
  AKT acknowledges support from the Natural Sciences and Engineering Research Council of Canada (NSERC) [PGSD3-545841-2020].
\end{acknowledgments}
\end{comment}

\bibliographystyle{IEEEtran}
\bibliography{refs}

\end{document}